\documentclass[twocolumn,amsmath,amssymb]{revtex4}

    \newenvironment{theorem}[1][Theorem]{\begin{trivlist}
    \item[\hskip \labelsep {\bfseries #1}]}{\end{trivlist}}
     \newenvironment{lemma}[1][Lemma]{\begin{trivlist}
    \item[\hskip \labelsep {\bfseries #1}]}{\end{trivlist}}

\def\PLA{{Phys. Lett.}  A}
\def\MPLA{{Mod. Phys. Lett.}  A}
\def\PRL{Phys. Rev. Lett.}

\def\PRB{{Phys. Rev.} B}
\def\PRA{{Phys. Rev.} A}

\def\JPA{{J. Phys.} A}

\def\JMP{{J. Math. Phys.}}

\begin{document}
\title{The challenge of non-hermitian structures in physics}
\author{A. Ram\'{\i}rez}
\email{aramirez@fis.cinvestav.mx}
\author{B. Mielnik }
\email{bogdan@fis.cinvestav.mx}
\affiliation{Departamento de F\'{\i}sica, CINVESTAV  A.P. 14-740, 07000 M\'exico, D.F.}

\begin{abstract}
        We present a brief review of physical problems leading to indefinite
	Hilbert spaces and non-hermitian Hamiltonians. With the exception of
	pseudo-Riemannian manifolds in GR, the problem of a consistent physical
	interpretation of these structures still waits to be faced.
\end{abstract}

\maketitle

\section{Introduction}
	Among attempts to generalize the orthodox quantum theory, a visible 
	place belongs to the indefinite metric and non-hermitian Hamiltonians. 
	The indefinite Hilbert spaces were proposed in 1942 by Dirac 
	\cite{dirac}. In 1950 Gupta applies the idea in QED to avoid the 
	negative energy photons \cite{gupt}. The pseudo-euclidean structures 
	are generic in relativistic theories. The mathematical studies of 
	Pontriagin \cite{pont}, Kre\u{\i}n et al. \cite{krein1}, \cite{ioh1}, 
	\cite{miel}, have offered the canonical forms of the pseudo-hermitian 
	operators (see also \cite{thomp1}, \cite{book}). A notable physical 
	case is the classification of the real, $4\times 4$ pseudo-hermitian 
	energy momentum tensor $T^{\alpha}_{\beta}$ in terms of the 
	``null-legs'' \cite{7}.

	The non-hermitian Hamiltonians (in genuine Hilbert spaces) appear in 
	problems of  the continuous measurements \cite{gis}, \cite{daniel},
	\cite{geneve} and ``time operator'' \cite{BJ}, in some works on 
	superconductors and other applied studies \cite{super1}, \cite{max1}, 
	\cite{max2}. As to the indefinite Hilbert spaces, they  awake a 
	comprehensible distrust due to the ``ghosts'' of negative probabilities 
	(in some occasions, though, enjoying better press than the negative 
	energies \cite{wig}, \cite{feyn}!). On the other hand, Bender et al. 
	\cite{bender1}-\cite{bender7} consider the non-hermitian 
	Hamiltonians with real spectra, enjoying the $\cal{PT}$ symmetry: 
	$H^\dagger=\mathcal{PT}H\mathcal{PT}$; the aspect of indefinite 
	metric being studied by Znojil \cite{znojil1}.
	In all these attempts the challenge of consistent physical 
	interpretation persists \cite{kocik}. To illustrate it, we quote 
	briefly the main structural facts concerning the pseudo-hermitian 
	operators in indefinite Hilbert spaces.
\section{Pseudo-hermitian operators}
	Let $X$ be a complex linear space, $dim\; X=N<+\infty $.
	The mapping $X\ni x,y \rightarrow \langle x,y\rangle \in \mathbb{C}$ 
	is called an  {\it indefinite scalar product} if:
	\begin{align}
	i.   &\quad \langle x,\alpha_1y_1+\alpha_2y_2\rangle =\alpha\langle 
	     x,y_1\rangle+\alpha_2\langle x,y_2\rangle\nonumber \\
	ii.  &\quad \langle x,y\rangle=\langle y,x\rangle^* \nonumber\\
	iii. &\quad \langle x,y\rangle= 0\;\;\forall x\;\;\;\Rightarrow\;\;\; 
	     y=0 \nonumber\\
	iv.  &\quad \exists\; x\neq 0 \text{ such that } \langle x,x\rangle=0 
	     \label{2.1}
	\end{align}
	
	The space $X$ with an indefinite scalar product is called {\it
	pseudo-euclidean}, or {\it indefinite Hilbert space}.
	
	A vector $x\in X$ such that $\langle x,x\rangle=0$ is called a {\it 
	null vector}. A vector $x\in X$ is called {\it positive} ({\it 
	negative}) if $\langle x,x \rangle \, >0\,(<0)$.
	
	Two vectors $x_1,x_2 \in X$ are called {\it orthogonal} if
	$\langle x_1,x_2 \rangle =0$. Every null vector is orthogonal to 
	itself. Two subspaces  $Z_1,Z_2\subset X$ are {\em orthogonal} if
	$\langle z_1,z_2 \rangle =0$ for every $z_1\in Z_1$ and $z_2\in Z_2$.
	A subspace $V\subset X$ such that $\langle u,v \rangle =0$ for all 
	$u,v\in V$ is called a {\it null subspace}.
	
	The subspace $Y\subset X$ is called {\it positive} ({\it negative}) if
	(i) $\langle x,x \rangle \;\geq 0\;(\leq 0)$ for every $x\in Y$, 
	(ii) $x\in Y$ and $\langle x,x \rangle =0$ implies $x=0$. In all 
	orthogonal decompositions $X=X_+\oplus X_-$ into the positive and 
	negative subspaces $X_+$ and $X_-$, the numbers $k=dimX_+$ and 
	$l=dimX_-$ are  the same. The pair $(k,l)$, $k+l=N$ corresponds to the 
	numbers of positive and negative vectors in any orthogonal basis and 
	is called the {\em signature} of $X$.
	
	Notice that the subspace $V\subset X$ is a null subspace iff $V$ is 
	orthogonal to itself. The dimensions of null subspaces in $X$ are 
	limited by the following 
	\begin{lemma}{\bf 1} (Pontriagin)
	The maximal possible number of linearly independent, mutually orthogonal
	null vectors in an indefinite Hilbert space of signature $(k,l)$ is 
	$r=min(k,l)$.
	\end{lemma}

	A subspace $Y\subset X$ is called {\it nonsingular} if $Y\cap
	Y^{\perp}=\{ 0\}$ and {\it singular} if the intersection $Y\cap
	Y^{\perp}$ contains at least one non-zero vector.	
	In particular, every positive (negative) subspace is nonsingular. 
	The whole $X$ is nonsingular due to ($1) iii$. 

	For every linear operator $A$ defined in $X$ there exists exactly one
	operator $A^{\dagger}$ such that
	\begin{equation}
	\langle x,Ay \rangle =\langle A^{\dagger} x,y\rangle\label{3.1}
	\end{equation}
	
	The operator $A$ is called {\it hermitian with respect to} 
	$\langle\,,\,\rangle$ (or {\em pseudo-hermitian}) if $A^\dagger =A$.
        For an indefinite product $\langle\,,\,\rangle$, several important 
	properties of the traditional self-adjoint operators do not hold:
	($a$) the pseudo-hermitian operators can have complex eigenvalues; 
	($b$) they do not need to be diagonalizable; 
	($c$) the eigenvectors are not necessarily orthogonal.
	All this can happen  due to the existence of non-trivial null vectors
	and nilpotent operators in $X$. The easiest examples illustrating  
	($a$)-($c$), are constructed with the help of the {\it dyadic 
	operators} $u\otimes v$ $=$ $|u\rangle\langle v|$ defined by
	\begin{equation}
	(u\otimes v)x=\langle v, x\rangle u\;\;\;\;\;\;\;\;\;\;(x\in X)\label{3.3} 
	\end{equation}
	
	Obviously, $(u\otimes v)^\dagger =v\otimes u$. Choosing now two null 
	vectors $v,\tilde {v}\in X$ with $\langle v,\tilde{v}\rangle=1$ and 
	putting
	\begin{equation}
	A=v\otimes \tilde{v} \quad \textrm{and}  \quad
	B= i (\tilde{v}\otimes v-v\otimes \tilde{v}) \label{3.6}
	\end{equation}
	one has $A^\dagger=A$, $B^\dagger=B$. Yet, $A^2=0$ though $A\neq 0$; 
	hence, $A$ is non-diagonalizable. Moreover, $Bv=-i v$ and $B\tilde{v}=i 
	\tilde{v}$; hence $B$ possesses the imaginary eigenvalues for two non
	orthogonal eigenvectors $v$ and $\tilde{v}$. As one can show, $A$ and 
	$B$ are the simplest bricks of the non-trivial  pseudo-hermitian 
	structures. In fact, the  Jordan's cell of any real eigenvalue 
	$\lambda$ of the hermitian $A=A^\dagger$ hosts a nilpotent hermitian operator 
	$Q=A-\lambda$, $Q^q=0$, $Q^{q-1}\neq 0$. If $2s\leq q$, the last $s$ 
	vectors in the Jordan's chain $x, Qx, ..., Q^{q-1}x$ must be  null and 
	mutually orthogonal (indeed, $i+j\geq q \Rightarrow\; \langle 
	Q^ix,Q^jx\rangle=\langle x,Q^{i+j}x\rangle =0$). The indefinite metric 
	permits the existence of such chains, though the Pontriagin's lemma 
	restricts their length to $s\leq r$. By choosing adequately the initial 
	vector $x$ one reduces the entire chain either to a  ``null leg''
	$v_1,...,v_s,\tilde{v}_1,...,\tilde{v}_s$ (where $\langle v_i,v_j
	\rangle =\langle \tilde{v}_i,\tilde{v}_j\rangle =0,\langle v_i,
	\tilde{v}_j\rangle =\pm\delta_{ij}$) or to a null leg plus a unit 
	vector $e$ ($\langle e,v_i\rangle =\langle e,\tilde{v}_i\rangle =0$, 
	$|\langle e,e\rangle |=1$) $v_1,...,v_s,e,\tilde{v}_s,...,\tilde{v}_1$; 
        a basis which brings $Q$ to one of the operational schemes:
	\begin{equation}
	v_1\;\rightarrow\;v_2\;\rightarrow\;\;\cdots\;\rightarrow\;v_s\;
	\rightarrow\;\tilde{v}_s
	\;\rightarrow\;\cdots\;\rightarrow\;\tilde{v}_1\;\rightarrow\;0
         \nonumber
	\end{equation}	
	\begin{equation}
	v_1\;\rightarrow\;\cdots\;\rightarrow\;v_s\;\rightarrow\;\;e\;\;
	\rightarrow\;\tilde{v}_s
	\;\rightarrow\;\cdots\;\rightarrow\;\tilde{v}_1\;\rightarrow\;0
         \nonumber
	\end{equation}
	with obvious canonical forms:
	{ \small\begin{eqnarray}	
	\hspace*{-1.5cm}
	\pm Q=
	v_2\otimes\tilde{v}_1\;\;+\;\;v_3\otimes\tilde{v}_2\;\;+\;\;\cdots
	\;\;+\;\;
	v_s\otimes\tilde{v}_{s-1} &&	\nonumber\\
	 \tilde{v}_s\otimes\tilde{v}_s\;+\; 	
	\tilde{v}_1\otimes v_2\;\;+\;\;\cdots\;\;+\;\;
	\tilde{v}_{s-1}\otimes v_s &\hspace*{.3cm}&	\label{4.3} \\
        &&\nonumber\\
	\pm Q=
	v_2\otimes\tilde{v}_1\;\;+\;\;\cdots\;\;+\;\;
	v_s\otimes\tilde{v}_{s-1}\;\;+\;\;e \otimes\tilde{v}_s	&&\nonumber\\
	+\;\;\tilde{v}_s \otimes e\;\;+	
	\tilde{v}_1\otimes v_2 \;\;+\;\;\cdots\;\;+\;\;
	\tilde{v}_{s-1}\otimes v_s	 &&		\label{4.4}
	\end{eqnarray}}
	
	The null legs are crucial as well for the  complex eigenvalues. Indeed, 
	if $A^\dagger =A$, each complex eigenvalue $\lambda=\alpha +i\beta$ is 
	accompanied by  $\lambda^*$ of equal multiplicity. Instead of analyzing 
	separately the Jordan's subspaces of $\lambda$ and $\lambda^*$ it is 
	more convenient to determine the structure of $A$ in their sum  
	$X_{(\lambda\;\lambda^*)}=X_{(\lambda)}+X_{(\lambda^*)}$ by using the 
	nilpotent pair $Q=A-\lambda$, $Q^\dagger=A-\lambda^*$. Since $Q$ and 
	$Q^\dagger$ are both nilpotent in $X_{(\lambda\;\lambda^*)}$, so is 
	$QQ^\dagger$. As one can  show, there must exists a vector $x\in  
	X_{(\lambda \;\lambda^*)}$ such that the triangle $\Delta$ of vectors
	{\footnotesize	
	\begin{equation} 
	\hspace*{-.5cm}	\begin{array}{ccccccccccc}
				&&&&&x&&&&& \\
			&&&&  Qx   & &Q^{\dagger}x&&&&\\
		&&& Q^2x & & QQ^{\dagger}x& & Q^{\dagger 2}x &&&\\
	\end{array}   \label{3.35}
	\end{equation} 
\begin{equation} 
\hspace*{-.5cm}	\begin{array}{ccccccccccc}
&	  && \cdots	&  	& \cdots &	&   \cdots    	&    & 	& \\
&	  &	& 	&  	& 	 &	&    	& 	& 	& \\
&&
Q^{2s-1}x&\cdots&Q^sQ^{\dagger s-1}x&&Q^{s-1}Q^{\dagger s}x&\cdots&Q^{\dagger 2s-1}x
&&
	\end{array}  \nonumber
	\end{equation} 
	}spans a nonsingular subspace  $X_\Delta \subset X_{(\lambda \;\lambda^*)}$, 
	where $Q^{2s}$ and $Q^{\dagger 2s}$ vanish but not $Q^{2s-1}$ and 
	$Q^{\dagger 2s-1}$. 
	The $2s$ vectors of the last row are linearly independent 
	and form a natural basis in the triangle subspace $X_\Delta$. Now, it 
	is easy to show that by choosing properly  the top vector $x\in 
	X_\Delta$ one can  reduce the basic row of (\ref{3.35}) to a null leg 
	defined as 
	{\footnotesize 
	\begin{equation}
	\begin{array}{cccccc}
	Q^{2s-1}x,&\cdots,&Q^sQ^{\dagger s-1}x, 
	& Q^{s-1}Q^{\dagger s}x,&\cdots,& Q^{\dagger{2s-1}}x \\
	\shortparallel &  & \shortparallel & \shortparallel &  &
	\shortparallel \\
	\tilde{v}_1 &  & \tilde{v}_s & v_1 & & v_s
	\end{array} \label{3.48}
	\end{equation}}
	
         Observing the action of $Q$ and $Q^\dagger$ on (\ref{3.48}) and
	remembering that $Q -Q^\dagger =2i\beta$, $Q^{2s}=Q^{\dagger 2s}=0$ 
         in $X_\Delta$, one sees that $A$ in $X_\Delta$ has the canonical form: 
	{\small	
	\begin{eqnarray}
	A_\Delta & = &\alpha\sum_{j=1}^s\big(\tilde{v}_j\otimes v_j + 
	v_j\otimes\tilde{v}_j\big)\; +\nonumber\\
	&& i\beta\sum_{j=1}^s\big(\tilde{v}_j\otimes v_j -
	v_j\otimes\tilde{v}_j\big)+  \nonumber\\
	&& 2i\beta\sum_{j=1}^{s-1}\sum_{i=j+1}^s
	\big(\tilde{v}_j\otimes v_i - v_i\otimes\tilde{v}_j\big)  \label{4.19}
	\end{eqnarray}}
         where $2s=\,dim X_\Delta\;\leq 2r$ due to lemma 1. As one can easily show, 
         each $X_{(\lambda\;\lambda^*)}$  decomposes into an orthogonal sum of 
	nonsingular
         triangular subspaces $X_\Delta$ where $A$ acquires the form (\ref{4.19}),
         so we have
	\begin{theorem}
        The hermitian operator $A$ in a pseudo-euclidean space $X$ is reducible
        to the  sequence of Jordan's  cells corresponding to  real 
         eigenvalues $\lambda_i$ where $Q_i=A-\lambda_i$ are of the canonical form 
	(\ref{4.3}-\ref{4.4}), and to a number of cells of  the complex 
	$\lambda_i$ $\lambda_i^*$ 
	with $A$ given by (\ref{4.19}), 
	the total number of  mutually orthogonal null vectors in all 
	irreducible cells of type (\ref{4.3}), (\ref{4.4}), (\ref{4.19}) being limited by the Pontriagin criterion.
	\end{theorem}   

\section{The physical problems}

	The pseudo-hermitian structures have a well defined status in
	classical theories. Thus, e.g., the energy momentum tensor 
	$T^{\alpha}_{\beta}$ of GR is  an example of a hermitian operator in 
	the pseudo-euclidean space of signature $(+ - -\;-)$. So, according to 
	the canonical forms of Sec. II, it can adopt only 4 basic types of
	Plebanski \cite{7}:
	(1) $[Z\!-\!Z^*\!\!-\!S_1\!-\!S_2]$ (diagonalizable, real eigenvalues
		    $S_1$, $S_2$, complex ones $Z$, $Z^*$);
	(2) $[T\!-\!S_1\!-\!S_2\!-\!S_3]$ (complete diagonalization with
		    four linearly independent eigenvectors);
	(3) $[2N\!-\!S_1\!-\!S_2]$ (non-diagonalizable canonical form 
		    which admits three eigenvectors);
	(4) $[3N\!-\!S_1]$ (non-diagonalizable canonical form admitting 
	only two eigenvectors).
	The analogous types would exist in more dimensions for the signature 
	$(1,l)$.

	An interesting case of {\em strings} living in a pseudo-euclidean space 
	$\mathbb{R}^4$ of signature $(+ + -\, -)$ is discussed in \cite{string}. 
	Notice that the field theories formulated in such space would lead to
	the energy momentum tensors of new algebraic types. Generalizing \cite{7}, 
	all of them can be reduced to the following standard forms:

	(1) $[2Z\!-\!2Z^*]$; two complex eigenvalues with nontrivial Jordan's cells;
	the canonical form 
	    $
	    A=\alpha [(\tilde{v}_1\otimes v_1+v_1\otimes\tilde{v}_1)+
	      (\tilde{v}_2\otimes v_2+v_2\otimes\tilde{v}_2)]+
            i\beta [(\tilde{v}_1\otimes v_1-v_1\otimes\tilde{v_1})+
    	      (\tilde{v}_2\otimes v_2-v_2\otimes\tilde{v_2})] +
            2i\beta(\tilde{v}_1\otimes v_2-v_2\otimes\tilde{v_1})	            
	    $;
	    
	(2) $[Z_1\!-\!Z_1^*\!-\!Z_2\!-\!Z_2^*]$, two pairs of complex conjugate
	eigenvalues, complete diagonalization:
	    $
	    A=\alpha_1(\tilde{v}_1\otimes v_1+v_1\otimes\tilde{v}_1)+  
	    i\beta_1(\tilde{v}_1\otimes v_1-v_1\otimes\tilde{v_1}) +
	    \alpha_2(\tilde{v}_2\otimes v_2+v_2\otimes\tilde{v}_2) + 
	    i\beta_2(\tilde{v}_2\otimes v_2-v_2\otimes\tilde{v_2});
	    $
	    
	(3) $[Z_1\!-\!Z_1^*\!-\!2S_1]$ with
	    $
	    A=\alpha_1(\tilde{v}_1\otimes v_1+v_1\otimes\tilde{v}_1) +
	    i\beta_1(\tilde{v}_1\otimes v_1-v_1\otimes\tilde{v_1})\pm
	    (\tilde{v}_2\otimes\tilde{v}_2)\pm\lambda_2(v_2\otimes\tilde{v}_2+
	     \tilde{v}_2\otimes v_2);  
	    $
	    
	(4) $[Z_1\!-\!Z_1^*\!-\!S_1\!-\!S_2]$   with
	    $
	    A=\alpha_1(\tilde{v}_1\otimes v_1+v_1\otimes\tilde{v}_1) + 
    	    i\beta_1(\tilde{v}_1\otimes v_1-v_1\otimes\tilde{v_1})\pm 
	    \lambda_2(e_2\otimes e_2)\pm\lambda_3(e_3\otimes e_3); 
	    $
	    
	(5) $[T_1\!-\!T_2\!-\!S_1\!-\!S_2]$ with the trivial
	    $
	    A=\lambda_1(e_1\otimes e_1)+
	      \lambda_2(e_2\otimes e_2)-
	      \lambda_3(e_3\otimes e_3)-
	      \lambda_4(e_4\otimes e_4) 
	    $
	    
	(6) $[2N_1\!-\!2N_2]$ with
	    $
	    A=\pm(\tilde{v}_1\otimes\tilde{v}_1)\pm\lambda_1(v_1\otimes\tilde{v}_1+
	     \tilde{v}_1\otimes v_1) \pm
	    (\tilde{v}_2\otimes\tilde{v}_2)\pm\lambda_2(v_2\otimes\tilde{v}_2+
	     \tilde{v}_2\otimes v_2);
	    $
	    
	(7) $[2N\!-\!S_1\!-\!S_2]$  with
	    $
	    A=\pm(\tilde{v}_1\otimes\tilde{v}_1)\pm\lambda_1(v_1\otimes\tilde{v}_1+
	     \tilde{v}_1\otimes v_1)+
	    \lambda_2(e_2\otimes e_2)+
	    \lambda_3(e_3\otimes e_3);  
	    $
	    
	(8) $[3N\!-\!S_1]$ with
	    $
	    A=\pm(e\otimes\tilde{v}_1+\tilde{v}_1\otimes e)\mp\lambda_1
	    (v_1\otimes\tilde{v}_1+\tilde{v}_1\otimes v_1+e\otimes e)
	    \mp\lambda_2(e_2\otimes e_2); 
	    $ 
	    
	(9) finally $[4N]$ with
	    $
	    A=\pm(v_2\otimes\tilde{v}_1+\tilde{v}_2\otimes\tilde{v}_2+
	    \tilde{v_1}\otimes v_2)\pm\lambda(v_1\otimes\tilde{v}_1+
	     \tilde{v}_1\otimes v_1+v_2\otimes\tilde{v}_2+
	     \tilde{v}_2\otimes v_2). 
	    $
	    
	(where $|\langle e_i,e_j\rangle|=|\langle v_i,\tilde{v}_j\rangle|=
	\delta_{ij}$, $\langle v_i,e\rangle=\langle \tilde{v}_i,e\rangle=
	\langle v_i,v_j\rangle=\langle \tilde{v}_i,\tilde{v}_j\rangle=0$).		    
	The question of physical nature of the corresponding sources remains 
	open.

	In quantum theories the main challenge is caused by negative vectors 
	(``ghosts''), as well as by the absence of a consistent measurement 
	theory. Thus, e.g. in an interesting plasma study (\cite{max1}) it is 
	assumed that the eigenvectors of the pseudo-hermitian operator must 
	form a basis  in $X$. However, it is not so: As we have seen, even in 
	the signature $(1,l)$ the subspaces of real eigenvalues may have a 
	nontrivial Jordan structure.

	In an ample class of quantum field theories following Gupta and Dirac  
	\cite{dirac}, \cite{gupt} the indefinite Hilbert spaces (with $dim X=+\infty$)
	 arise 
	as an auxiliary element,  eliminated later  by the constraints conditions 
	to avoid the ``ghost'' vectors. The original field operators turn as well 
	``ghost observables''  to be substituted by the constrained ones 
	\cite{strocchi2}, \cite{strocchi1}, \cite{dimock}. However, once all 
	ghosts depart, an unsolved mystery remains why the  ghost formulation 
	of the theory was at all necessary?

	The situation is different in several areas of QM  showing a 
	`non-hermitian dissidence' which cannot be maintained on purely ghost 
	level. Notice the fundamental role of non-hermitian Hamiltonians (in 
	the  orthodox Hilbert spaces) for the continuous reduction processes 
	\cite{gis}, \cite{daniel}, \cite{geneve}, \cite{BJ}. For different 
	reasons the complex Hamiltonians with real spectra are studied in 
	\cite{bender1}-\cite{bender7}. Here, the story develops on the 
	heuristic level of $\mathcal{PT}-$symmetric operators. In mathematical 
	terms, it can be viewed as a new branch of the spectral analysis in 
	Banach spaces. However the link with the pseudo-euclidean structures  
	was recently reported by Znojil \cite{znojil1}. Independent steps in the 
	same direction are taken by Takook \cite{ren1}. In all these designs the 
	consistent statistical interpretation is still missing. So, will the 
	indefinite Hilbert spaces contribute only to a new ``ghost story'' or 
	are they a real escape route from too much orthodoxy in  quantum theory?

	The authors acknowledge the stimulating discussion of Lech Woronowicz, 
	Carl Bender, Miloslav Znojil during the QTS conference in Krakow in 
	July 2001 and of Daniel Sudarsky  and Oscar Rosas on the Workshop in 
	Chapala, Mexico; November 2001. The support of Conacyt project $32086-$E 
	is acknowledged.



\begin{thebibliography}{99}
\bibitem{dirac} Dirac, P.A.M.
         Proc Roy Soc. A  $\;\mathbf{180}$ 1 (1942).
\bibitem{gupt} Gupta, S. N.  
	Proc. Phys. Soc. Sect. A.  $\;\mathbf{63}$, 681 (1950).
\bibitem{pont} Pontriagin, L.S. 
	Izv. Akad. Nauk. SSSR Ser Mat  $\;\mathbf{8}$, 243 (1944).
\bibitem{krein1} Kre\u{\i}n, M. G. and Rutman, M.A.
	Amer. Math. Soc. Transl. $\;\mathbf{26}$ 199 (1950). 
\bibitem{ioh1} Iohvidov, I.S. and Kre\u{\i}n, M. G.
         Amer. Math. Soc. Transl. Series 2.  $\;\mathbf{13}$ 105, (1960);  
         $\;\mathbf{34}$  283 (1963).
\bibitem{miel} Mielnik, B. {\em Hermitean matrices in spaces with indefinite scalar product}
	Report CINVESTAV  (1965).
\bibitem{thomp1} Thompson, R.C.
         Linear algebra Appl. {\bf 14}, 135 (1976) and  {\bf 147}, 323 (1991).
\bibitem{book} Gohberg I., Lancaster P., Rodman L.
         {\em Matrices and Indefinite Scalar Products}, Birkh\"auser, Basel, (1983).
\bibitem{7} Plebanski, J. 
	Acta Phys. Pol. XXVI 963 (1964).
\bibitem{gis} Gisin, N., J.Phys.A {\bf 14}, 2259 (1981)
\bibitem{daniel} Daniel, W., Helv.Phys.Acta {\bf 55},
330 (1982)                
\bibitem{geneve} Huttner, B; Muller, A; Gautier, J.D.; Zbinden, H. and Gisin N.
\PRA  $\;\mathbf{54}$ 3783 (1996).
\bibitem{BJ} Blaunchard, Ph. and Jadczyk, A. Helv Phys. Acta  $\;\mathbf{69}$ 
613 (1996). 
\bibitem{super1} Hatano, N. and Nelson, D. R.
     \PRB $\;\mathbf{56}$, 8651 (1997); 
     $\;\mathbf{77}$,  570 (1996).
\bibitem{max1} Larsson J.
     \PRL $\;\mathbf{66}$, 1466 (1991).   \\
      
\bibitem{max2} Brizard, A.J., Cook D.R., Kaufman A.N.
     \PRL $\;\mathbf{70}$, 521 (1993).      
\bibitem{wig} Wigner E.P.  
     Phys. Rev. {\bf 40}, 749 (1932).
\bibitem{feyn} Feynman  in: Hiley and Peat eds. Quantum Implications. Routledge and
     Keagan, London 235 (1987).
\bibitem{bender1} Bender, C., Boettcher, S.
     \JPA $\;\mathbf{31}$, L273 (1998);
     \PRL $\;\mathbf{80}$,  5243 (1998). 
\bibitem{bender3} Bender, C., Boettcher, S., Meisinger, P.N.
     \JMP $\;\mathbf{40}$, 2210 (1999).
\bibitem{bender5} Bender, C.,  Dunne,G. V.
     \JMP $\;\mathbf{40}$, 4616 (1999)   
\bibitem{bender6} Bender, C.,  Dunne,G. V.C., Meisinger, P.N.
     \PLA $\;\mathbf{252}$, 272 (1999)
\bibitem{bender7} Bender, C.,  Milton, K.A., Meisinger, P.N.
     \JMP $\;\mathbf{40}$, 2201 (1999).         
\bibitem{znojil1} Znojil, M. 
     math-ph/0106021,
     \PLA   $\;\mathbf{285}$, 7 (2001), 
     math-ph/0104012,
     math-ph/0103054.    
\bibitem{kocik} Kocik, J.
     Int. J. Theor. Phys.  $\;\mathbf{38}$ 2221 (1999).
\bibitem{string} Ooguri, H.
     \MPLA  $\;\mathbf{5}$ 1389 (1990).
\bibitem{strocchi2}  Strocchi, F. Wightman, A. S. 
     \JMP  $\;\mathbf{15}$  2198 (1974).      
\bibitem{strocchi1} Morchio, G. Strocchi, F.
     Ann. Inst. Henri Poincar\'e. A  {\bf XXXIII}  251 (1980).      
\bibitem{dimock} Dimock, J. math-ph0102027
\bibitem{ren1} Takook, M. V.
      gr-qc/0006052;  gr-qc/0006019.          
\end{thebibliography}
\end{document}